\pdfoutput=1
\documentclass[bib]{statapress}

\usepackage[crop,newcenter,frame]{pagedims}

\usepackage{bm} 
\usepackage{bbm}
\usepackage{amsmath}
\usepackage{amssymb}
\usepackage{amsfonts}
\usepackage{subcaption}
\usepackage{float}
\usepackage{enumitem}
\usepackage{multirow}
\usepackage{tabularx}
\usepackage{xspace}
\usepackage{xcolor}
\usepackage{booktabs}
\usepackage{setspace}
\usepackage[normalem]{ulem} 
\usepackage{enumitem}
\usepackage[flushleft]{threeparttable}  
\usepackage{url}
\usepackage{fancyvrb}
\VerbatimFootnotes

\usepackage{sj}
\usepackage{epsfig}
\usepackage{stata}
\usepackage{shadow}
\sjsetissue{$vv$}{$ii$}{$mm$}{$yyyy$}

\usepackage{accents}
\newlength{\dhatheight}

\begin{document}


\graphicspath{{fig/}}

\makeatletter
\def\input@path{{tab/}{mc/}}
\makeatother


\newcommand{\sjversion}[2]{#1}

\newcommand{\norm}[1]{\left\lVert#1\right\rVert}
\newcommand{\normline}[1]{\lVert#1\rVert}

\newcommand{\myoption}[2]{\texttt{#1(}#2\texttt{)}}
\newcommand{\ttt}[1]{\texttt{#1}}
\newcommand{\lassopack}{\stcmd{lassopack}\xspace}
\newcommand{\pdslasso}{\stcmd{pdslasso}\xspace}
\newcommand{\lassotwo}{\stcmd{lasso2}\xspace}
\newcommand{\cvlasso}{\stcmd{cvlasso}\xspace}
\newcommand{\rlasso}{\stcmd{rlasso}\xspace}
\newcommand{\lasso}{lasso\xspace}
\newcommand{\ddml}{\stcmd{ddml}\xspace}
\newcommand{\qddml}{\stcmd{qddml}\xspace}
\newcommand{\pystacked}{\stcmd{pystacked}\xspace}
\newcommand{\pylasso}{\stcmd{pylasso}\xspace}
\newcommand{\sklearn}{{\it scikit-learn}\xspace}

\newcommand{\real}{{\it real}}
\newcommand{\stint}{{\it integer}}
\newcommand{\stvarname}{{\it varname}}
\newcommand{\var}{{\it variable}}
\newcommand{\method}{{\it method}}
\newcommand{\stnumlist}{{\it numlist}}
\newcommand{\stmatrix}{{\it matrix}}

\newcommand{\stopt}[2]{\texttt{#1(}\emph{#2}\texttt{)}}

\newcommand{\achim}[1]{\textcolor{red}{$<<$AA: #1$>>$}}
\newcommand{\ms}[1]{\textcolor{blue}{$<<$MS: #1$>>$}}
\newcommand{\chris}[1]{\textcolor{violet}{$<<$CH: #1$>>$}}

\newcommand{\aarevised}[2]{\textcolor{red}{\sout{#1} #2}}
\newcommand{\msrevised}[2]{\textcolor{blue}{\sout{#1} #2}}


\inserttype[st0001]{article}
\author{Ahrens, Hansen \& Schaffer}{%
  Achim Ahrens\\ETH Z\"urich \\achim.ahrens@gess.ethz.ch
  \and
  Christian B. Hansen\\  University of Chicago\\christian.hansen@chicagobooth.edu
  \vspace{.3cm}
  \and
 Mark E. Schaffer\\ Heriot-Watt University\\Edinburgh, United Kingdom \\ m.e.schaffer@hw.ac.uk
}
\title[pystacked]{pystacked: Stacking generalization and machine learning in Stata}

\maketitle
\sjversion{}{\thispagestyle{empty}}


\begin{abstract}
\pystacked implements stacked generalization (Wolpert, 1992) for regression and binary classification via Python's \emph{scikit-learn}. Stacking combines multiple supervised machine learners---the ``base'' or ``level-0'' learners---into a single learner. The currently supported base learners include regularized regression, random forest, gradient boosted trees, support vector machines, and feed-forward neural nets (multi-layer perceptron). \pystacked can also be used as a `regular' machine learning program to fit a single base learner and, thus, provides an easy-to-use API for \emph{scikit-learn}'s machine learning algorithms.

\end{abstract}


\section{Introduction}
When faced with a new prediction or classification task, it is \emph{a priori} rarely obvious which machine learning algorithm is best suited. A common approach is to evaluate the performance of a set of machine learners on a hold-out partition of the data or via cross-validation and then select the machine learner that minimizes a chosen loss metric. However, this approach is incomplete as combining multiple learners into one final prediction might lead to superior performance compared to each individual learner. This possibility motivates stacked generalization, or simply \emph{stacking}, due to \citet{Wolpert1992} and \citet{Breiman1996a}. Stacking is a form of model averaging. Theoretical results in \citet{Laan2007} support the use of stacking as it performs asymptotically at least as well as the best-performing individual learner as long as the number of base learners is not too large.

In this article, we introduce \pystacked for stacking regression and binary classification in Stata. \pystacked allows users to fit multiple machine learning algorithms via Python's \sklearn \citep{scikit-learn,sklearn_api} and combine these into one final prediction as a weighted average of individual predictions. \pystacked adds to the growing number of programs for machine learning in Stata, including, among others, \texttt{lassopack} for regularized regression \citep{Ahrens2019}, \texttt{rforest} for random forests \citep{Schonlau2020} and \texttt{svm} for support vector machines \citep{Schonlau2016,Guenther2018}. Similarly to \pystacked, \citet{cerulli2022machine} and \citet{Droste2020} provide an interface to \sklearn in Stata. \texttt{mlrtime} allows Stata users to make use of R's \emph{parsnip} machine learning library \citep{Huntington2021}. \pystacked differs from these in that it is, to our knowledge, the first to make stacking available to Stata users. Furthermore, \pystacked can also be used to fit a single machine learner and thus provides an easy-to-use and versatile API to \sklearn's machine learning algorithms.

Stacking is widely used in applied predictive modeling in many disciplines, e.g., for predicting mortality \citep{hwangbo2022}, bankruptcy filings \citep{liang2020,fedorova2022}, or temperatures \citep{hooker2018}. The use of stacking as a method and \pystacked as a program is, however, not only restricted to pure prediction or classification tasks. A growing literature  exploits machine learning to facilitate causal inference \citep[see for an overview][]{Athey2019c}. Indeed, a motivation for writing \pystacked is that it can be used in combination with \ddml \citep{ddml_package,ddml_papers}, which implements the Double-Debiased Machine Learning (DDML) methodology of \citet{Chernozhukov2018}. DDML utilizes cross-fitting, a form of iterative sample splitting, which allows leveraging a wide class of supervised machine learners, including stacking, for the estimation of causal parameters. For instance, in the context of DDML, stacking can be used to estimate the conditional expectation of an outcome with respect to confounders or propensity scores.

We stress that \pystacked relies on Python's \sklearn (version 0.24 or higher) and the ongoing work of the \sklearn contributors. Thus, \pystacked relies on Stata's Python integration which was introduced in Stata 16.0.  We kindly ask users to cite \sklearn along with this article when using \pystacked. Throughout we refer to version 0.7 of \pystacked, which is the latest version at the time of writing. Please check for updates to \pystacked on a regular basis and consult the help file to be informed about new features. The \pystacked help file includes information on how to install a recent Python version and set up Stata's Python integration.

Section~\ref{sec:method} introduces stacking. Section~\ref{sec:program} presents the main features of the \pystacked program. Section~\ref{sec:applications} demonstrates the use of the program using examples.
    
\section{Methodology}\label{sec:method}
In this section, we briefly summarize the stacking approach for regression and binary classification tasks. \citet{van2011targeted} provides a book-length treatment; for a concise textbook treatment, see \citet{Hastie2009}. 

We first focus on stacking for regression problems where the aim is to predict the continuous outcome $y_i$ using predictors $\bm{x}_i$.
The idea of stacking is to combine a set of ``base'' (or ``level-0'') learners using a ``final'' (or ``level-1'') estimator. It is advisable to include a relatively large and diverse set of base learners to capture different types of patterns in the data. The same algorithm can also be included more than once using different tuning or hyper-tuning parameters. Typical choices for base learners are regularized regression or ensemble methods, such as random forests or gradient boosting. 

In the first step of stacking, we obtain cross-validated predicted values $\hat{y}^{-k(i)}_{(j),i}$ for each base learner $j$ and observation $i$. The super-script ``$-k(i)$'' indicates that we form the cross-fitted predicted value for observation $i$ by fitting the learner to all folds except fold $k(i)$, which is the fold that includes observation $i$. The use of cross-validation is necessary as stacking would otherwise give more weight to base learners that suffer from over-fitting. The second step is to fit a final learner using the observed $y_i$ as the outcome and the cross-validated predicted values $\hat{y}^{-k(i)}_{(1),i},\ldots,\hat{y}^{-k(i)}_{(J),i}$ as predictors. A typical choice for the final learner is constrained least squares, which enforces the stacking weights to be non-negative and sum to one. This restriction facilitates the interpretation of stacking as a weighted average of base learners and may lead to better performance \citep{Breiman1996a,Hastie2009}. Algorithm~1 summarizes the stacking algorithm for regression problems as it is implemented in \pystacked. 

\begin{sttech}[Algorithm 1: Stacking regression.]
\begin{enumerate}[nosep] 
    \item \emph{Cross-validation:}
    \begin{enumerate}
        \item Split the sample $I=\{1,\ldots,n\}$ randomly into $K$ partitions of approximately equal size. These partitions are referred to as \emph{folds}.  Denote the set of observations in fold $k=1,\ldots,K$ as $I_k$, and its complement as $I_k^c$ such that $I_k^c=I \setminus I_k$. $I_k$ constitutes the step-$k$ validation set and $I_k^c$ the step-$k$ training sample.
        \item For each fold $k=1,\ldots,K$ and each base learner $j=1,\ldots,J$, fit the supervised machine learner $j$ to the training data $I_k^c$ and obtain out-of-sample predicted values $\hat{y}^{-k}_{(j),i}$ for $i\in I_k$.
    \end{enumerate}
    \item \emph{Final learner:} Fit the final learner to the full sample. The default choice is non-negative least squares (NNLS) with the additional constraint that coefficients sum to one:
    \[ \underset{w_1,\ldots,w_J}{\min} \sum_{i=1}^n \left(y_i - \sum_{j=1}^Jw_j\hat{y}^{-k(i)}_{(j),i}\right)^2 \qquad \textrm{s.t.}\quad w_j\geq 0,~~~ \sum_{j=1}^Jw_j=1\]
    The stacking predicted values are defined as $\hat{y}^\star_i=\sum_j\hat{w}_j\hat{y}_{(j),i}$ where $\hat{w}_j$ is the estimated stacking weight corresponding to learner $j$ and $\hat{y}_{(j),i}$ are the predicted values from re-fitting learner $j$ on the full sample $I$.
\end{enumerate}
\end{sttech}

It is instructive to compare Step~2 with the classical `winner-takes-all' approach that selects one base learner as the one which exhibits the lowest cross-validated loss. Stacking, in contrast, may assign non-zero weights to multiple base learners, thus combining their strengths to produce a better overall predictor than any of the individual base learners. Other choices for the final learner are possible. In addition to the default final learner, \pystacked supports, among others, non-negative least squares without the constraint $\sum w_j=1$, ridge regression, the aforementioned `winner-takes-all' approach that selects the base learner with the smallest cross-validated mean-squared error, and unconstrained least squares.

\subsection{Stacking classification}
Stacking can be applied in a similar way to classification problems. \pystacked supports stacking for binary classification problems where the outcome $y_i$ takes the values 0 or~1. The main difference to stacking regression is that  $\hat{y}^{-k}_{(j),i}$ represent cross-validated predicted probabilities. 

\subsection{Base learners}
In the following paragraphs, we briefly describe the base learners supported by \pystacked and highlight central tuning parameters. We repeat that each of the machine learners discussed below can be fit using \pystacked as a regular stand-alone machine learner without the stacking layer.

Note that it goes beyond the scope of the article to describe each learner in detail. Familiarity with linear regression, logistic regression, and classification and regression trees is assumed. We recommend consulting machine learning textbooks, e.g.\ \citet{Hastie2009}, for more detailed discussion. 

\paragraph{Regularized regression} imposes a penalty on the size of coefficients to control over-fitting. The lasso penalizes the absolute size of coefficients, whereas the ridge penalizes the sum of squared coefficients. Both methods shrink coefficients toward zero, but only the lasso yields sparse solutions where some coefficient estimates are set to exactly zero. The elastic net combines lasso and ridge-type penalties. For classification tasks with a binary outcome, logistic versions of lasso, ridge and elastic net are available.

The severity of the penalty is most commonly chosen by cross-validation. For lasso only, \pystacked also supports selecting the penalty by AIC or BIC. The use of AIC or BIC has the advantage that it is computationally less intensive than cross-validation. \citet{Ahrens2019} compare the two approaches.

\paragraph{Random forests} rely on fitting a large number of regression or decision trees on bootstrap samples of the data. The random forest prediction is obtained as the average across individual trees. A crucial aspect of random forests is that, at each split when growing a tree, one may consider only a random subset of predictors. This restriction aims at de-correlating the individual trees. Central tuning parameters are the number of trees (\verb+n_estimators()+), the maximum depth of individual trees (\verb+max_depth()+), the minimum number of observations per leaf (\verb+min_samples_leaf()+), the number of features to be considered at each split (\verb+max_features()+) and the size of the bootstrap samples (\verb+max_samples()+).

\paragraph{Gradient boosted trees} also rely on fitting a large number of trees. In contrast to random forests, these trees are fit sequentially to the residuals from the current model. The learning rate determines how much the latest tree contributes to the overall model. Individual trees are usually fit to the whole sample, although sub-sampling is possible. In addition to tuning parameters relating to the trees, the learning rate (\verb+learnings_rate()+) and the number of trees (\verb+n_estimators()+) are the most important tuning parameters.

\paragraph{Support vector machines (SVM).} Support vector classifiers span a hyperplane that separates observations by their outcome class. The hyperplane is chosen to maximize the distance (\emph{margin}) to correctly classified observations while allowing for some classification errors. The tuning parameter $C$ (\verb+C()+) controls the frequency and degree of classification mistakes. The hyperplane can be either linear or fitted using kernels.
The SVM algorithm can also be adapted for regression tasks. To this end, the hyperplane is constructed to include as many observations as possible in a tube of size $2\epsilon$ around the hyperplane. Central tuning parameters for regression are $\epsilon$ (\verb+epsilon()+) and $C$ (\verb+C()+), which determines the cost of observations outside of the tube.
  
\paragraph{Feed-forward neural networks} consist of hidden layers that link the predictors (referred to as \emph{input layers}) to the outcome. Each hidden layer is composed of multiple units (\emph{nodes}) which pass signals to the next layer using an activation function. Central tuning parameters are the choice of the activation function (\verb+activation()+), and the number and size of hidden layers (\verb+hidden_layer_sizes()+). Further tuning choices relate to stochastic gradient descent algorithms which are typically used to fit neural networks. The default solver is Adam \citep{kingma2014adam}. The option \verb+early_stopping+ can be used to set aside a random fraction of the data for validation. The optimization algorithm stops if there is no improvement in performance over a pre-specified number of iterations (see related options \verb+n_iter_no_change()+ and \verb+tol()+).

\section{Program}\label{sec:program}
This section introduces the program \pystacked and its main features. \pystacked offers two alternative syntaxes between which the user can choose (see Section~\ref{sec:syntax1} and \ref{sec:syntax2}). The two syntaxes offer the same functionality and are included to accommodate different user preferences. Section~\ref{sec:postest} and \ref{sec:other_options} list post-estimation commands and general options, respectively. Section~\ref{sec:base_learners} discusses supported base learners. Section~\ref{sec:predictors} is a note on learner-specific predictors. Section~\ref{sec:pipelines} explains the pipeline feature.

\subsection{Syntax 1}\label{sec:syntax1}
The first syntax uses {\tt\underbar{m}ethod}(\ststring) to select base learners, where {\it string} is a list of base learners.  Options are passed on to base learners via \stopt{cmdopt1}{string}, \stopt{cmdopt2}{string}, etc. That is, base learners can be specified and options are passed on in the order in which they appear in {\tt\underbar{m}ethod}(\ststring) (see Section~\ref{sec:base_learners}).  Likewise, the 
\stopt{pipe*}{string} option can be used for pre-processing predictors within Python on the fly, where `\texttt{*}' is a placeholder for `\texttt{1}', `\texttt{2}', etc. (see Section~\ref{sec:pipelines}). Finally, \stopt{xvars*}{predictors} allows specifying a learner-specific variable lists of predictors.

\begin{stsyntax}
pystacked
    \depvar\
    {\it predictors}\
    \optif\
    \optin\
    \optional{,
    \underbar{m}ethods(\ststring)
    cmdopt1(\ststring)
    cmdopt2(\ststring)
    ...
    pipe1(\ststring)
    pipe2(\ststring)
    ...
    xvars1({\it predictors})
    xvars2({\it predictors})
    ...
    {\it general\_options}
    }
\end{stsyntax}

\noindent where {\it general\_options} are discussed in Section~\ref{sec:other_options} below.

\subsection{Syntax 2}\label{sec:syntax2}
In the second syntax, base learners are added before the comma using {\tt\underbar{m}ethod}(\ststring) along with further learner-specific settings and separated by `\verb+||+'.
    
\begin{stsyntax}
pystacked
    \depvar\
    \optindepvars\
    \verb+||+ \textnormal{{\tt\underbar{m}ethod}(\ststring) 
    {\tt opt}(\ststring)
    {\tt pipe}(\ststring)
    {\tt xvars}({\it predictors})}
    \optional{|| \underbar{m}ethod(\ststring) 
    opt(\ststring)
    pipe(\ststring)
    xvars({\it predictors})
    ... ||}
    \optif\
    \optin\
    \optional{,
    {\it general\_options}
    }
\end{stsyntax}

\subsection{Post-estimation programs}\label{sec:postest}

\paragraph{Predicted values.}
To get predicted values:
\begin{stsyntax}
 predict \emph{type} {\it newname} \optif\ \optin\ \optional{, pr xb }
 \end{stsyntax}

To get fitted values for each base learner:
\begin{stsyntax}
 predict \emph{type} {\it stub} \optif\ \optin\   \optional{, \underbar{base}xb \underbar{cv}alid }
\end{stsyntax}

Predicted values (in- and out-of-sample) are calculated when \pystacked is run and stored in
Python memory. \texttt{predict} pulls the predicted values from Python memory and saves them in Stata memory. This storage structure means that no changes on the data in Stata memory should be made between the \pystacked call and the \texttt{predict} call. If changes to the data set are made, \texttt{predict} will return an error.

The option {\tt \underline{base}xb} returns predicted values for each base learner. By default, the predicted values from re-fitting base learners on the full estimation sample
are returned. If combined with {\tt \underline{cv}alid}, the cross-fitted predicted values are returned for each base learner.

\paragraph{Tables.}
After estimation, \pystacked can report a table of in-sample (both cross-validated and full-sample) and, optionally, out-of-sample (or holdout sample) performance for both the stacking regression and the base learners.  For regression  problems, the table reports the root mean-squared prediction error (RMSPE). For classification problems, a confusion matrix is reported.  The default holdout sample used for out-of-sample performance with the holdout option is all observations not included in the estimation.  Alternatively, the user can specify the holdout sample explicitly using the syntax \stopt{holdout}{varname}. The table can be requested after estimation as a replay command or as part of the \pystacked estimation.


\begin{stsyntax}
pystacked  \optional{, \underline{tab}le holdout[({\it varname})] }
\end{stsyntax}

\paragraph{Graphs.}
\pystacked can also create graphs of in-sample and, optionally, out-of-sample performance for both the stacking regression and the base learners. For regression problems, the graphs compare predicted and actual values of \emph{depvar}. For classification problems, the default is to generate receiver operator characteristic (ROC) curves. Optionally, histograms of predicted probabilities are reported. As with the {\tt table} option, the default holdout sample used for out-of-sample performance is all observations not included in the estimation, but the user can instead specify the holdout sample explicitly.  The table can be requested after estimation or as part of the \pystacked estimation command. The {\tt graph} option on its own reports the graphs using \pystacked's default settings.  Because graphs are produced using Stata's \texttt{twoway}, \texttt{roctab} and \texttt{histogram} commands, the user can control either the combined graph (\stopt{graph}{options}) or the individual learner graphs (\stopt{lgraph}{options}) by passing options to these commands.
    
\begin{stsyntax}
 pystacked  \optional{, graph[({\it options})] lgraph[({\it options})] \underline{hist}ogram holdout[({\it varname})] }
\end{stsyntax}

\subsection{General options}\label{sec:other_options}
A full list of general options is provided in the \pystacked help file. We list only the most important general options here:

\begin{description}[noitemsep]
    \item[\stopt{\uline{ty}pe}{{\it string}}] allows \emph{reg(ress)} for regression problems or \emph{class(ify)} for classification problems. The default is regression.
    \item[\stopt{\underline{final}est}{{\it string}}]  selects the final estimator used to combine base learners. The default is non-negative least squares without an intercept and the additional constraint that weights sum to 1 ({\it nnls1}). Alternatives are {\it nnls0} (non-negative least squares without an intercept and without the sum-to-one constraint), {\it singlebest} (use the base learner with the minimum MSE), {\it ls1} (least squares without an intercept and with the sum-to-one constraint), {\it ols} (ordinary least squares) or {\it ridge} for (logistic) ridge, which is the \sklearn default. 
    \item[\stopt{\underline{f}olds}{{\it integer}}] specifies the number of folds used for cross-validation. The default is 5.
    \item[\stopt{\underline{foldv}ar}{{\it varname}}] is the integer fold variable for cross-validation.
    \item[\stopt{\underline{bf}olds}{{\it integer}}] sets the number of folds used for base learners that use cross-validation (e.g. cross-validated lasso); the default is 5.
    \item[\stopt{\uline{pys}eed}{{\it integer}}] sets the Python seed. Note that since \pystacked uses Python, we also need to set the Python seed to ensure replicability. There are three options: \begin{enumerate} \item \stopt{pyseed}{-1} draws a number between 0 and $10^8$ in Stata which is then used as a Python seed; this is \pystacked's default behavior. This way, one only needs to deal with the Stata seed. For example, {\tt set seed 42} is sufficient, as the Python seed is generated automatically. \item Setting \stopt{pyseed}{x} with any positive integer \emph{x} allows to control the Python seed directly. \item \stopt{pyseed}{0} sets the seed to \texttt{None} in Python. \end{enumerate}
    \item[\stopt{\uline{nj}obs}{{\it integer}}] sets the number of jobs for parallel computing. The default is 0 (no parallelization), --1 uses all available CPUs, --2 uses all CPUs minus 1. 
    \item[\stopt{backend}{{\it string}}] 
    backend used for parallelization. The default is `threading'.
    \item[{\tt \underline{vot}ing}] selects voting regression or classification which uses pre-specified weights. By default, voting regression uses equal weights; voting classification uses a majority rule. 
    \item[\stopt{\underline{votew}eights}{{\it numlist}}] defines positive weights used for voting regression or classification. The length of {\it numlist} should be the number of base learners -- 1. The last weight is calculated to ensure that the sum of weights equals 1.
    \item[{\tt \uline{sp}arse}] converts predictor matrix to a sparse matrix. This conversion will only lead to speed improvements if the predictor matrix is sufficiently sparse. Not all learners support sparse matrices and not all learners will benefit from sparse matrices in the same way. One can also use the sparse pipeline to use sparse matrices for some learners but not for others.
    \item[{\tt \uline{print}opt}]
    prints the default options for specified learners.
    Only one learner can be specified.
    This is for information only;
    no estimation is done. See Section~\ref{sec:base_learners}
    for examples.
    \item[{\tt \uline{show}opt}]
    prints the options passed on to Python.
    \item[{\tt \uline{showp}y}]
    prints Python messages.
    \end{description}

\subsection{Base learners}\label{sec:base_learners}
The base learners are chosen using the option \stopt{\uline{m}ethod}{string} in combination with \stopt{type}{string}. The latter can take the value \emph{reg(ress)} for regression and \emph{class} for classification problems. Table~\ref{tab:base_learners} provides an overview of supported base learners and their underlying \sklearn routines.

\stopt{cmdopt*}{string} (Syntax~1) and \stopt{opt}{string} (Syntax~2) are used to pass options to the base learners. Due to the large number of options, we do not list all options here. We instead provide a tool that lists options for each base learner. For example, to get the default options for lasso with cross-validated penalty, type

\begin{minipage}{\linewidth}
\begin{stlog}
. pystacked, type(reg) methods(lassocv) printopt
Default options: 
alphas(None) l1_ratio(1) eps(.001) n_alphas(100) fit_intercept(True)
max_iter(1000) tol(.0001) cv(5) n_jobs(None) positive(False) selection(cyclic)
random_state(rng) 
\end{stlog}
\vspace{.2cm}
\end{minipage}

The naming of the options follows \sklearn. Allowed settings for each option can be inferred from the \sklearn documentation. We strongly recommend that the user reads the \sklearn documentation carefully.\footnote{The \sklearn documentation is available at \url{https://scikit-learn.org/stable/index.html} (last accessed on August 18, 2022).} 
    
\begin{table}
    \centering\footnotesize
    \begin{threeparttable}
    \begin{tabularx}{1\linewidth}{>{\it}l>{\it}l>{\raggedright}X>{\tt}l}
    \hline\hline
    {\textnormal{\tt method()}} & {\textnormal{\tt type()}}      & \emph{Machine learner description} & \emph{scikit-learn program} \\\hline
      ols                   & regress & Linear regression &  linear\_model.LinearRegression\\
      logit                 & class & Logistic regression  & linear\_model.LogisticRegression \\
      lassoic               & regress & Lasso with AIC/BIC penalty& linear\_model.LassoLarsIC  \\
      lassocv               & regress & Lasso with CV penalty & linear\_model.ElasticNetCV \\
                            & class & Logistic lasso with CV penalty & linear\_model.LogisticRegressionCV \\
      ridgecv               & regress & Ridge with CV penalty & linear\_model.ElasticNetCV \\
                            & class & Logistic ridge with CV penalty & linear\_model.LogisticRegressionCV \\
      elasticcv             & regress & Elastic net with CV penalty& linear\_model.ElasticNetCV \\
                            & class & Logistic elastic net with CV& linear\_model.LogisticRegressionCV\\
      svm                   & regress & Support vector regression & svm.SVR\\
                            & class & Support vector classification & svm.SVC \\
      gradboost             & regress & Gradient boosting regression &ensemble.GradientBoostingRegressor  \\
                            & class   & Gradient boosting classification & ensemble.GradientBoostingClassifier\\
      rf                    & regress & Random forest regression & ensemble.RandomForestRegressor \\    
                            & class & Random forest classification & ensemble.RandomForestClassifier \\
      linsvm               
                            & class & Linear SVC & svm.LinearSVC \\
      nnet                  & regress & Neural net regression & sklearn.neural\_network.MLPRegressor\\
                            & class & Neural net classification &sklearn.neural\_network.MLPClassifier \\ 
    \hline\hline\\[-.3cm]
    \end{tabularx}
    \begin{tablenotes}
    \item \emph{Note:} The first two columns list all allowed combinations of \stopt{method}{string} and \stopt{type}{string}, which are used to select base learners. Column 3 provides a description of each machine learner. The last column lists the underlying \sklearn learn routine. `CV penalty' indicates that the penalty level is chosen to minimize the cross-validated MSPE. `AIC/BIC penalty' indicates that the penalty level minimizes either either the Akaike or Bayesian information criterion. SVC refers to support vector classification.
    \end{tablenotes}
    \end{threeparttable}
    \caption{Overview of machine learners available in \pystacked.}
    \label{tab:base_learners}
\end{table}    

\subsection{Learner-specific predictors}\label{sec:predictors}
By default, \pystacked uses the same set of predictors for each base learner. Using the same predictors for each method is often not desirable as the optimal set of predictors may vary across base learners. For example, when using linear machine learners such as the lasso, adding polynomials, interactions and other transformations of the base set of predictors might greatly improve out-of-sample prediction performance. The inclusion of transformations of base predictors is especially worth considering if the base set of observed predictors is small (relative to the sample size) and the relationship between outcome and predictors is likely non-linear. Tree-based methods (e.g., random forests and boosted trees), on the other hand, can detect certain types of non-linear patterns automatically. While adding transformations of the base predictors may still lead to performance gains, the added benefit is less striking relative to linear learners and might not justify the additional costs in terms of computational complexity. 

There are two approaches to implement learner-specific sets of predictors: Pipelines, discussed in the next section, can be used to create some transformations on the fly for specific base learners. A more flexible approach is the \stopt{xvars*}{predictors} option, which allows specifying predictors for a particular learner. \stopt{xvars*}{} supports standard Stata factor variable notation.

\subsection{Pipelines}\label{sec:pipelines}
\sklearn uses pipelines to pre-preprocess input data on the fly. In \pystacked, pipelines can be used to impute missing values, create polynomials and interactions, and to standardize predictors. Table~\ref{tab:pipelines} lists the pipelines currently supported by \pystacked.    

\begin{table}
    \centering\footnotesize
    \begin{tabularx}{\linewidth}{>{\it}l>{\raggedright}X>{\tt}l}\hline\hline
      {\textnormal{\tt pipe*()}} &\emph{Description} & {\textnormal{\it scikit-learn programs}} \\
      \hline
      stdscaler           & Standardize to mean zero and unit variance (default for regularized regression) & StandardScaler()\\
      nostdscaler           & Overwrites default standardization for regularized regression & \textnormal{n/a} \\
      stdscaler0          & Standardize to unit variance  & StandardScaler(with\_mean=False)\\
      sparse              & Transform to sparse matrix & SparseTransformer()\\
      onehot              & Create dummies from categorical variables & OneHotEncoder() \\
      minmaxscaler        & Scale to 0-1 range & MinMaxScaler()\\
      medianimputer       & Median imputation & SimpleImputer(strategy='median')\\
      knnimputer          & KNN imputation & KNNImputer() \\
      poly2               & Add 2nd-order polynomials & PolynomialFeatures(degree=2) \\
      poly3               & Add 3rd-order polynomials & PolynomialFeatures(degree=3) \\
      interact            & Add interactions & PolynomialFeatures( \\ && ~~include\_bias=False, \\ && ~~interaction\_only=True) \\
      \hline\hline
    \end{tabularx}
    \caption{Pipelines supported by \pystacked.}
    \label{tab:pipelines}
\end{table}
 
\paragraph{Remarks.} First, regularized regressors (i.e., the methods {\it lassoic}, {\it lassocv}, {\it ridgecv} and {\it elasticcv}) use the \emph{stdscaler} pipeline by default. \stopt{pipe*}{nostdscaler} disables this behavior. Second, the \emph{stdscaler0} pipeline is useful in combination with \emph{sparse}, which transforms the predictor matrix into a sparse matrix. \emph{stdscaler0} does not center predictors so that the predictor matrix retains its sparsity property. 

\section{Applications}\label{sec:applications}
This section demonstrates how to apply \pystacked for regression and classification tasks. Before we discuss stacking, we first show how to use \pystacked as a `regular' machine learning program for fitting a single supervised machine learner (see next sub-section). We then illustrate stacking regression and stacking classification in Section~\ref{sec:appl_reg} and \ref{sec:appl_class}, respectively.

\subsection{Single base learner}\label{sec:single_learner}
We import the \citet{Kelley1997} California house price data, and split the sample randomly into training and validation partitions using a 75/25 split. The aim of the prediction task is to predict median house prices ({\tt medhousevalue}) using a set of house price characteristics.%
\footnote{The following predictors are included in the data set in this order:
district longitude (\texttt{longitude}), latitude (\texttt{latitude}), median house age (\texttt{houseage}), average number of rooms per household (\texttt{rooms}), average number of bedrooms per household (\texttt{bedrooms}), block group population (\texttt{population}), average number of household members (\texttt{households}), median income in block group (\texttt{medinc}).
The data set was retrieved from the StatLib repository at \verb|https://www.dcc.fc.up.pt/~ltorgo/Regression/cal_housing.html| (last accessed on February 19, 2022).
There are 20,640 observations in the data set.
} We prepare the data for analysis as follows.

\begin{stlog}
. clear all
{\smallskip}
. use https://statalasso.github.io/dta/cal_housing.dta, clear
{\smallskip}
. set seed 42
{\smallskip}
. gen train=runiform()
{\smallskip}
. replace train=train<.75
(20,640 real changes made)
{\smallskip}
. replace medh = medh/10e3 
variable {\bftt{medhousevalue}} was {\bftt{long}} now {\bftt{double}}
(20,640 real changes made)

\end{stlog}

\paragraph{Gradient-boosted trees.} As a first example, we use \pystacked to fit gradient-boosted regression trees and save the out-of-sample predicted values. Since we consider a regression task rather than a classification task, we specify \texttt{type(reg)} (which is also the default). The option \texttt{method(gradboost)} selects gradient boosting. We will later see that we can specify more than one learner in \stopt{methods}{}, and that we can also fit gradient-boosted classification trees. 

\begin{stlog}
. pystacked medh longi-medi if train, type(reg) methods(gradboost)
Single base learner: no stacking done.
{\smallskip}
Stacking weights:
\HLI{17}{\TOPT}\HLI{21}
  Method         {\VBAR}      Weight
\HLI{17}{\PLUS}\HLI{21}
  gradboost      {\VBAR}      1.0000000
{\smallskip}
. predict double yhat_gb1 if !train
{\smallskip}

\end{stlog}

The output shows the stacking weights associated with each base learner. Since we only consider one method, the output is not particularly informative and simply shows a weight of one for gradient boosting. Yet, \pystacked has fitted 100 boosted trees (the default) in the background using \sklearn's \texttt{ensemble.}\-\texttt{Gradient}\-\texttt{Boosted}\-\texttt{Regressor}. 

Before we tune our gradient boosting learner, we retrieve a list of available options. The default options for gradient boosting can be listed in the console, and the \sklearn documentation provides more detail on the allowed parameters of each option. 

\begin{stlog}
. pystacked, type(reg) methods(gradboost) printopt
Default options: 
loss(squared_error) learning_rate(.1) n_estimators(100) subsample(1)
criterion(friedman_mse) min_samples_split(2) min_samples_leaf(1) 
min_weight_fraction_leaf(0) max_depth(3) min_impurity_decrease(0) init(None)
random_state(rng) max_features(None) alpha(.9) max_leaf_nodes(None)
warm_start(False) validation_fraction(.1) n_iter_no_change(None)
tol(.0001) ccp_alpha(0)
\end{stlog}

We consider two additional specifications. Note that we restrict ourselves to a few selected specifications for illustrative purposes. We stress that careful parameter tuning should consider a grid of values across multiple learner parameters. The first specification reduces the learning rate from 0.1 (the default) to 0.01. The second specification reduces the learning rate and increases the number of trees from 100 (the default) to 1000.  We use \stopt{cmdopt1}{} since gradient boosting is the first (and only) method listed in \stopt{methods}{}. 

\begin{stlog}
. pystacked medh longi-medi if train,                        ///
>     type(regress) methods(gradboost)                       ///
>     cmdopt1(learning_rate(.01)) 
Single base learner: no stacking done.
{\smallskip}
Stacking weights:
\HLI{17}{\TOPT}\HLI{21}
  Method         {\VBAR}      Weight
\HLI{17}{\PLUS}\HLI{21}
  gradboost      {\VBAR}      1.0000000
{\smallskip}
. predict double yhat_gb2 if !train
{\smallskip}
. pystacked medh longi-medi if train,                        ///
>     type(regress) methods(gradboost)                       ///
>     cmdopt1(learning_rate(.01) n_estimators(1000)) 
Single base learner: no stacking done.
{\smallskip}
Stacking weights:
\HLI{17}{\TOPT}\HLI{21}
  Method         {\VBAR}      Weight
\HLI{17}{\PLUS}\HLI{21}
  gradboost      {\VBAR}      1.0000000
{\smallskip}
. predict double yhat_gb3 if !train
{\smallskip}

\end{stlog}

We can then compare the performance across the three models using the out-of-sample mean-squared prediction error (MSPE):

\begin{stlog}
. gen double res_gb1_sq=(medh-yhat_gb1){\caret}2 if !train
(15,448 missing values generated)
{\smallskip}
. gen double res_gb2_sq=(medh-yhat_gb2){\caret}2 if !train
(15,448 missing values generated)
{\smallskip}
. gen double res_gb3_sq=(medh-yhat_gb3){\caret}2 if !train
(15,448 missing values generated)
{\smallskip}
. sum res_gb* if !train
{\smallskip}
    Variable {\VBAR}        Obs        Mean    Std. dev.       Min        Max
\HLI{13}{\PLUS}\HLI{57}
  res_gb1_sq {\VBAR}      5,192    29.86727    83.17099   3.92e-06   1424.438
  res_gb2_sq {\VBAR}      5,192    71.10136    129.1446   4.56e-07   1230.348
  res_gb3_sq {\VBAR}      5,192    30.36929    82.88687   1.24e-06   1384.924
{\smallskip}

\end{stlog}

The initial gradient booster achieves an out-of-sample MSPE of 29.87. The second gradient booster uses a reduced learning rate of 0.01 and performs much worse, with an MSPE of 71.10. The third gradient booster performs only slightly worse than the first, illustrating the trade-off between the learning rate and the number of trees.

\paragraph{Pipelines.} We can make use of pipelines to pre-process our predictors. This is especially useful in the context of stacking when we want to, for example, use second-order polynomials of predictors as inputs for one method, but only use elementary predictors for another method. Here, we compare lasso with and without the \emph{poly2} pipeline:

\begin{stlog}
. pystacked medh longi-medi if train, type(reg) methods(lassocv)  
Single base learner: no stacking done.
{\smallskip}
Stacking weights:
\HLI{17}{\TOPT}\HLI{21}
  Method         {\VBAR}      Weight
\HLI{17}{\PLUS}\HLI{21}
  lassocv        {\VBAR}      1.0000000
{\smallskip}
. predict double yhat_lasso1 if !train
{\smallskip}
. 
. pystacked medh longi-medi if train, type(reg) methods(lassocv) ///
>     pipe1(poly2)   
Single base learner: no stacking done.
{\smallskip}
Stacking weights:
\HLI{17}{\TOPT}\HLI{21}
  Method         {\VBAR}      Weight
\HLI{17}{\PLUS}\HLI{21}
  lassocv        {\VBAR}      1.0000000
{\smallskip}
. predict double yhat_lasso2 if !train
{\smallskip}

\end{stlog}

We could replace {\small \stopt{pipe1}{\tt poly2}} with \texttt{\small xvars1(c.(medh longi-medi)\#\#c.(medh longi -medi))}. In fact, the latter is more flexible and allows, for example, to create interactions for some predictors and not for others.

We again calculate the out-of-sample MSPE:

\begin{stlog}
. gen double res_lasso1_sq=(medh-yhat_lasso1){\caret}2 if !train
(15,448 missing values generated)
{\smallskip}
. gen double res_lasso2_sq=(medh-yhat_lasso2){\caret}2 if !train
(15,448 missing values generated)
{\smallskip}
. sum res_lasso1_sq res_lasso2_sq if !train
{\smallskip}
    Variable {\VBAR}        Obs        Mean    Std. dev.       Min        Max
\HLI{13}{\PLUS}\HLI{57}
res_lasso1{\tytilde}q {\VBAR}      5,192    47.00385    108.0759   8.43e-07   2392.572
res_lasso2{\tytilde}q {\VBAR}      5,192    43.81224    109.5096   6.75e-08    2563.15
{\smallskip}

\end{stlog}

The \emph{poly2} pipeline improves the performance of the lasso, indicating that squared and interaction terms constitute important predictors. However, the lasso does not perform as well as gradient boosting in this application.

\subsection{Stacking regression}\label{sec:appl_reg}

We now consider a stacking regression application with five base learners: (1) linear regression, (2) lasso with penalty chosen by cross-validation, (3) lasso with second-order polynomials and interactions, (4) random forest with default settings, and (5) gradient boosting with a learning rate of 0.01 and 1000 trees.
That is, we use the lasso twice---once with and once without the \emph{poly2} pipeline. Indeed, nothing keeps us from using the same algorithm multiple times. This way, we can combine the same algorithm with different settings.

Note the numbering of the \stopt{pipe*}{} and \stopt{cmdopt*}{} options below. We apply the \emph{poly2} pipe to the first and third methods (\emph{ols} and \emph{lassoic}). We also change the default learning rate and the number of estimators for gradient boosting (the 5th estimator).

\begin{stlog}
. set seed 42
{\smallskip}
. pystacked medh longi-medi if train,                        ///
>     type(regress)                                          ///
>     methods(ols lassocv lassocv rf gradboost)              ///
>     pipe3(poly2) cmdopt5(learning_rate(0.01)               ///
>     n_estimators(1000))
{\smallskip}
Stacking weights:
\HLI{17}{\TOPT}\HLI{21}
  Method         {\VBAR}      Weight
\HLI{17}{\PLUS}\HLI{21}
  ols            {\VBAR}      0.0000000
  lassocv        {\VBAR}      0.0000000
  lassocv        {\VBAR}      0.0000000
  rf             {\VBAR}      0.8382714
  gradboost      {\VBAR}      0.1617286
{\smallskip}

\end{stlog}

The above syntax becomes a bit difficult to read with many methods and many options. We offer an alternative syntax for easier use with many base learners. 

\begin{stlog}
. set seed 42
{\smallskip}
. pystacked medh longi-medi                                   || ///
>     m(ols)                                                  || ///
>     m(lassocv)                                              || ///
>     m(lassocv) pipe(poly2)                                  || ///
>     m(rf)                                                   || ///
>     m(gradboost) opt(learning_rate(0.01) n_estimators(1000))   ///
>     if train, type(regress) 
{\smallskip}
Stacking weights:
\HLI{17}{\TOPT}\HLI{21}
  Method         {\VBAR}      Weight
\HLI{17}{\PLUS}\HLI{21}
  ols            {\VBAR}      0.0000000
  lassocv        {\VBAR}      0.0000000
  lassocv        {\VBAR}      0.0000000
  rf             {\VBAR}      0.8382714
  gradboost      {\VBAR}      0.1617286
{\smallskip}

\end{stlog}


The stacking weights shown in the output determine how much each method contributes to the final stacking predictor. In this example, OLS and lasso based on both sets of predictors all receive a weight of zero. Random forest receives a weight of 83.8\%, and gradient boosting contributes the remaining 16.2\% of the weight to the final predictor.

\paragraph{Predicted values.} In addition to the stacking predicted values, we can also get the predicted values of each base learner using the {\tt basexb} option:

\begin{stlog}
. predict double yhat, xb
{\smallskip}
. predict double ybase, basexb
{\smallskip}
. list yhat ybase* if _n <= 5
{\smallskip}
       {\TLC}\HLI{71}{\TRC}
       {\VBAR}      yhat      ybase1      ybase2      ybase3      ybase4      ybase5 {\VBAR}
       {\LFTT}\HLI{71}{\RGTT}
    1. {\VBAR} 42.875233   41.315834   41.193416    40.02005    43.16083   41.394926 {\VBAR}
    2. {\VBAR}  38.73664    41.45306   41.421461   44.443346   38.289107   41.056291 {\VBAR}
    3. {\VBAR}  40.68331   38.212036   38.154834   37.796287   41.268902   37.648071 {\VBAR}
    4. {\VBAR} 33.559085   32.332498   32.266321   32.546423     33.4869   33.933232 {\VBAR}
    5. {\VBAR} 24.190615   25.382839   25.369064   25.269325     23.8597   25.905815 {\VBAR}
       {\BLC}\HLI{71}{\BRC}

\end{stlog}

\begin{figure}[htb]
\centering
\includegraphics[width=.9\linewidth]{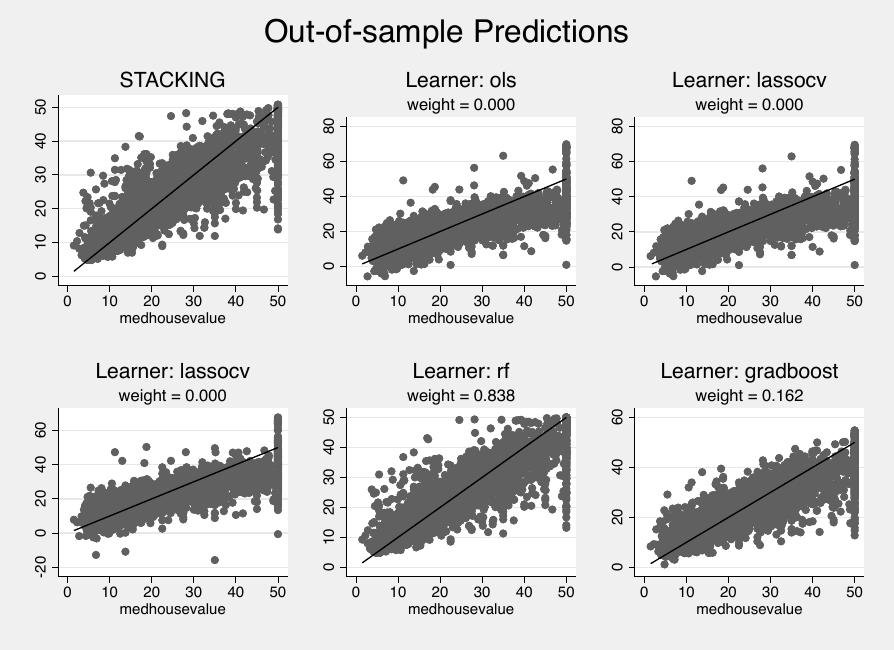}
\caption{Out-of-sample predicted values and observed values created using the \texttt{graph option} after stacking regression.}
\label{fig:stacking_reg}
\end{figure}

\paragraph{Plotting.} \pystacked also comes with plotting features. The \texttt{graph} option creates a scatter plot of predicted values on the vertical and observed values on the horizontal axis for stacking and each base learner, see Figure~\ref{fig:stacking_reg}. The black line is a 45-degree line and shown for reference. Since \pystacked with \texttt{graph} can be used as a post-estimation command, there is no need to re-run the stacking estimation. 

\begin{stlog}
. pystacked, graph(scheme(sj)) lgraph(scheme(sj)) holdout
Number of holdout observations: 5192
{\smallskip}

\end{stlog}

Figure~\ref{fig:stacking_reg} shows the out-of-sample predicted values. To see the in-sample predicted values, simply omit the \texttt{holdout} option. Note that the \texttt{holdout} option will not work if the estimation was run on the whole sample.

\hspace*{-\parindent}%
\begin{minipage}{\textwidth}
\paragraph{RMSPE table.} The \texttt{table} option allows comparing stacking weights with in-sample, cross-validated and out-of-sample RMSPEs. As with the \texttt{graph} option, we can use \texttt{table} as a post-estimation command:

\begin{stlog}
. pystacked, table holdout
Number of holdout observations: 5192
{\smallskip}
RMSPE: In-Sample, CV, Holdout
\HLI{17}{\TOPT}\HLI{47}
  Method         {\VBAR} Weight   In-Sample        CV         Holdout
\HLI{17}{\PLUS}\HLI{47}
  STACKING       {\VBAR}    .        2.313        4.980        4.939
  ols            {\VBAR} 0.000       6.986        7.008        6.853
  lassocv        {\VBAR} 0.000       6.987        7.008        6.857
  lassocv        {\VBAR} 0.000       6.696        6.699        6.606
  rf             {\VBAR} 0.838       1.847        5.001        4.963
  gradboost      {\VBAR} 0.162       5.312        5.523        5.511
{\smallskip}

\end{stlog}
\end{minipage}

\subsection{Stacking classification}\label{sec:appl_class}
\pystacked can be applied to binary classification problems. For demonstration, we consider the Spambase Data Set of \citet{cranor1998spam}, which we retrieve from the UCI Machine Learning Repository. We load the data and split the data into training (75\%) and validation sample (\%25).

\begin{stlog}
. insheet using ///
>  https://archive.ics.uci.edu/ml/machine-learning-databases/spambase/spambase.data, ///
>  clear comma
(58 vars, 4,601 obs)
{\smallskip}
. set seed 42
{\smallskip}
. gen train=runiform()
{\smallskip}
. replace train=train<.75
(4,601 real changes made)

\end{stlog}

The example below is more complicated. We go through it step-by-step:
\begin{itemize}[noitemsep]
    \item We use five base learners: logistic regression, two gradient boosters and two neural nets.
    \item  We apply the \emph{poly2} pipeline to logistic regression, which creates squares and interaction terms of the predictors, but not to other methods.
   \item We employ gradient boosting with 600 and 1000 classification trees. 
    \item We consider two specifications for the neural nets: one neural net with two hidden layers of 5 nodes each, and another neural net with a single hidden layer of 5 nodes.
    \item Finally, we use \stopt{type}{class} to specify that we consider a classification task and \stopt{njobs}{8} switches parallelization on utilizing 8 cores.
\end{itemize}

\begin{stlog}
. pystacked v58 v1-v57                            || ///
>     m(logit) pipe(poly2)                        || ///
>     m(gradboost) opt(n_estimators(600))         || ///
>     m(gradboost) opt(n_estimators(1000))        || ///
>     m(nnet) opt(hidden_layer_sizes(5 5))        || ///
>     m(nnet) opt(hidden_layer_sizes(5))          || ///
>     if train, type(class) njobs(8) 
{\smallskip}
Stacking weights:
\HLI{17}{\TOPT}\HLI{21}
  Method         {\VBAR}      Weight
\HLI{17}{\PLUS}\HLI{21}
  logit          {\VBAR}      0.0000000
  gradboost      {\VBAR}      0.4815155
  gradboost      {\VBAR}      0.3446655
  nnet           {\VBAR}      0.1131199
  nnet           {\VBAR}      0.0606992
{\smallskip}

\end{stlog}

As in the previous regression example, gradient boosting receives the largest stacking weight and thus contributes most to the final stacking prediction.

\paragraph{Confusion matrix.} 
Confusion matrices allow comparing actual and predicted outcomes in a $2\times 2$ matrix. \pystacked provides a compact table format that combines confusion matrices for each base learner and for the final stacking classifier, both for the training and validation partition.

\begin{minipage}{\linewidth}
\begin{stlog}
. pystacked, table holdout
Number of holdout observations: 1133
{\smallskip}
Confusion matrix: In-Sample, CV, Holdout
\HLI{17}{\TOPT}\HLI{59}
  Method         {\VBAR} Weight      In-Sample             CV             Holdout
                 {\VBAR}             0       1         0       1         0       1
\HLI{17}{\PLUS}\HLI{59}
  STACKING     0 {\VBAR}    .      2079       6      2004      81       678      29
  STACKING     1 {\VBAR}    .         2    1381        77    1306        29     397
  logit        0 {\VBAR} 0.000     1265      88      1231      86       413      32
  logit        1 {\VBAR} 0.000      816    1299       850    1301       294     394
  gradboost    0 {\VBAR} 0.482     2078       9      2004      83       679      30
  gradboost    1 {\VBAR} 0.482        3    1378        77    1304        28     396
  gradboost    0 {\VBAR} 0.345     2079       1      2002      86       680      29
  gradboost    1 {\VBAR} 0.345        2    1386        79    1301        27     397
  nnet         0 {\VBAR} 0.113     1834      55      1975     132       617      28
  nnet         1 {\VBAR} 0.113      247    1332       106    1255        90     398
  nnet         0 {\VBAR} 0.061     2005     146      1924     116       671      51
  nnet         1 {\VBAR} 0.061       76    1241       157    1271        36     375
{\smallskip}

\end{stlog}
\end{minipage}

For example, the table shows 29 false positives for stacking and 294 for logistic regression in the validation partition, while the number of false negatives is 29 and 32, respectively. The accuracy of stacking and logistic regression is thus given by $(678+397)/1133=94.9$\%  and $(413+394)/1133=71.2$\%.

\begin{figure}[htb]
\centering
\includegraphics[width=.9\linewidth]{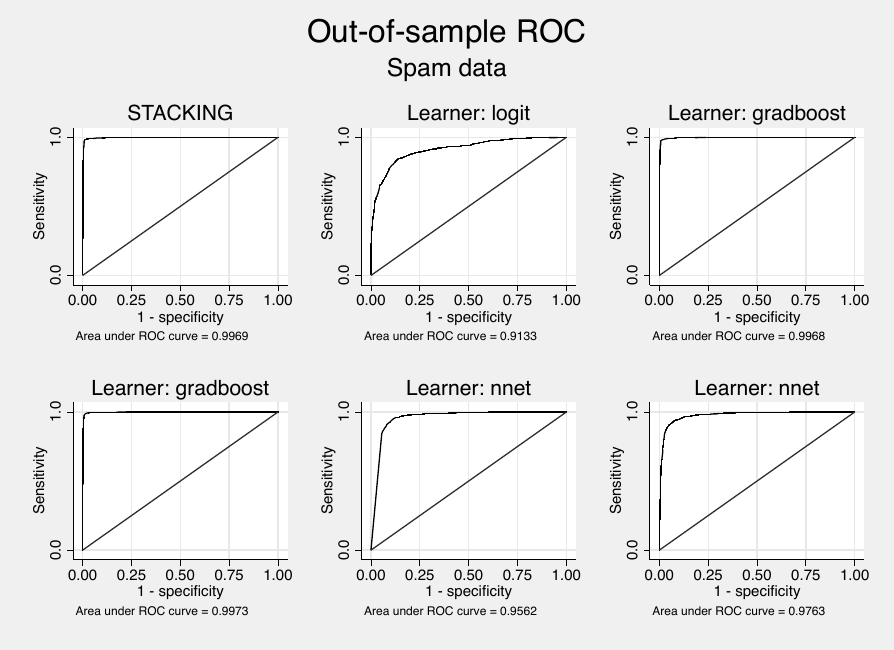}
\caption{Out-of-sample ROC curves created using the \texttt{graph option} after stacking regression.}
\label{fig:stacking_class}
\end{figure}

\paragraph{Plotting.} \pystacked supports ROC curves which allow assessing the classification performance for varying discrimination thresholds. The $y$-axis in a ROC plot corresponds to sensitivity (true positive rate) and the $x$-axis corresponds to 1--specificity (false positive rate). The Area Under the Curve (AUC) displayed below each ROC plot is a common evaluation metric for classification problems.

\begin{stlog}
. pystacked,                                  ///
>     graph(subtitle(Spam data) scheme(sj))   ///
>     lgraph(plotopts(msymbol(i)              ///
>     ylabel(0 1, format(\%3.1f))) scheme(sj)) ///
>     holdout
Number of holdout observations: 1133
{\smallskip}

\end{stlog}

\section{Conclusion}\label{sec:conclusion}
In this article, we introduce the Stata program \texttt{pystacked}. The program not only makes a range of popular supervised machine learners available, such as regularized regression, random forests, and gradient-boosted trees, but also facilitates combining multiple learners for stacking regression or classification. \texttt{pystacked} comes with a number of practically relevant features: for example, sparse matrix support, learner-specific predictors, pipelines for predictor transformations, and two alternative syntaxes. Nevertheless, we also see the potential for improvement in at least two directions. First, \texttt{pystacked} offers few diagnostics for individual learners. Features such as variable importance plots and coefficient estimates for parametric learners could be added in later versions. Secondly, the deep learning features of \texttt{pystacked} are currently relatively limited and could be improved by adding support for other deep learning algorithms.

\section{Acknowledgments}
We thank Jan Ditzen, Ben Jann, Blaise Melly, Alessandro Oliveira, Matthias Schonlau, and Thomas Wiemann for their helpful feedback. All remaining errors are our own. 

\bibliographystyle{sj}
\bibliography{library}

\begin{aboutauthors}
Achim Ahrens is Post-Doctoral Researcher and Senior Data Scientist at the Public Policy Group and Immigration Policy Lab, ETH Z\"urich.

Christian B. Hansen is the Wallace W. Booth Professor of Econometrics and Statistics at the University of Chicago Booth School of Business.

Mark E. Schaffer is Professor of Econonomics in the School of Social Sciences at Heriot-Watt University, Edinburgh, UK, and a Research Fellow at the Institute of Labor Economics (IZA), Bonn.
\end{aboutauthors}
\end{document}